\documentclass[twocolumn,english,aps,twocolumn]{revtex4}
\usepackage[T1]{fontenc}
\usepackage[latin9]{inputenc}
\usepackage{textcomp}
\usepackage{amsthm}
\usepackage{amsmath}
\usepackage{graphicx}
\usepackage{amssymb}
\usepackage{esint}

\makeatletter

\newcommand{\lyxmathsym}[1]{\ifmmode\begingroup\def\b@ld{bold}
  \text{\ifx\math@version\b@ld\bfseries\fi#1}\endgroup\else#1\fi}

\@ifundefined{textcolor}{}
{%
 \definecolor{BLACK}{gray}{0}
 \definecolor{WHITE}{gray}{1}
 \definecolor{RED}{rgb}{1,0,0}
 \definecolor{GREEN}{rgb}{0,1,0}
 \definecolor{BLUE}{rgb}{0,0,1}
 \definecolor{CYAN}{cmyk}{1,0,0,0}
 \definecolor{MAGENTA}{cmyk}{0,1,0,0}
 \definecolor{YELLOW}{cmyk}{0,0,1,0}
 }

\makeatother

\usepackage{babel}

\begin{document}

\title{Strong Coupling of a Spin Ensemble to a Superconducting Resonator}

\author{Y. Kubo$^{1}$, F. R. Ong$^{1}$, P. Bertet$^{1}$, D. Vion$^{1}$,
V. Jacques$^{2}$, D. Zheng$^{2}$, A. Dréau$^{2}$, J.-F. Roch$^{2}$,
A. Auffeves$^{3}$, F. Jelezko$^{4}$, J. Wrachtrup$^{4}$, M.F. Barthe$^{5}$,
P. Bergonzo$^{6}$, and D. Esteve$^{1}$}

\affiliation{$^{1}$Quantronics group, SPEC (CNRS URA 2464), IRAMIS, DSM, CEA,
91191 Gif-sur-Yvette, France }

\affiliation{$^{2}$LPQM (CNRS UMR 8537), ENS de Cachan, 94235 Cachan, France}

\affiliation{$^{3}$Institut Néel - CNRS-UJF 38042 Grenoble cedex 9, France }

\affiliation{$^{4}$3. Physikalisches Institut, Universität Stuttgart, 70550 Stuttgart,
Germany }

\affiliation{$^{5}$CNRS, UPR3079 CEMHTI, 1D avenue de la Recherche Scientifique,
45071 Orléans, France}

\affiliation{$^{6}$CEA, LIST, Diamond Sensors Laboratory, F-91191 Gif-sur-Yvette,
France}

\date{\today}
\begin{abstract}
We report the realization of a quantum circuit in which an ensemble
of electronic spins is coupled to a frequency tunable superconducting
resonator. The spins are nitrogen-vacancy centers in a diamond crystal.
The achievement of strong coupling is manifested by the appearance
of a vacuum Rabi splitting in the transmission spectrum of the resonator
when its frequency is tuned through the nitrogen-vacancy center electron
spin resonance.
\end{abstract}
\maketitle
Building a quantum computer requires engineering a system that can
reliably store quantum information and process it through a succession
of quantum gates. Quantum bit implementations based on individual
microscopic systems such as atoms, photons, electron or nuclear spins,
benefit from a natural decoupling from environmental noise, which
results in long coherence times \cite{WinelandRMP}; on the other
hand, superconducting qubits \cite{QubitsSupra} are macroscopic artificial
atoms that couple strongly to electromagnetic fields, allowing faster
single- and two-qubit gates, although with shorter coherence times
\cite{DiCarlo2qb}. It is thus appealing to take the best of both
worlds by combining artificial and natural quantum systems in {}``hybrid''
quantum circuits that would exhibit long coherence times while allowing
rapid quantum state manipulation. Whereas the coupling strength $g$
of an individual microscopic emitter to one electromagnetic mode is
usually too weak to allow for coherent exchange of quantum information,
the coupling strength of an ensemble of $N$ such systems is enhanced
by $\sqrt{N}$ \cite{VacuumRabiKimble,VacuumRabiRydberg,WallraffSqrtN},
allowing to reach the strong coupling regime $g\sqrt{N}\gg\kappa,\gamma$,
where $\kappa$ and $\gamma$ are the resonator and emitter damping
rates. In that perspective, it has been proposed to couple an ensemble
of cold molecules \cite{HybridsZoller}, atoms \cite{HybridsKurizki,HybridsSchmiedmayer},
or electron spins \cite{HybridsImamoglu,HybridsMolmer} to a superconducting
resonator that would mediate their interactions with one or a few
superconducting qubits. This system would work like a genuine quantum
Turing machine, with the ensemble providing a quantum memory \cite{LukinQMemory,GisinQMemory,BriggsEnsemble}
with long storage time, and the qubits providing the {}``hardware''
to perform quantum gates \cite{MolmerHolographicQC}.

Among all microscopic systems that can be coupled to superconducting
circuits, negatively charged nitrogen-vacancy centers (N-\emph{V})
in diamond are particularly attractive \cite{NVFluxQLukin}. They
consist of a substitutional nitrogen atom and an adjacent vacancy
having trapped an additional electron; their electronic ground state
has a spin $S=1$ with the states $m_{\mathrm{S}}=0$ and $m_{\mathrm{S}}=\pm1$
separated by $\sim2.87$~GHz in zero magnetic field \cite{NVJelezko}.
The coherence time corresponding to this transition has been demonstrated
to be as long as $2$\ ms at room-temperature for samples isotopically
enriched in $^{12}\mathrm{C}$ \cite{NVLongCoherence}. Compared to
atoms, N-\emph{V} centers are perfectly compatible with superconducting
circuits, because they do not require challenging trapping techniques
or large magnetic fields to bring them in resonance at GHz frequency
with the circuit. Finally, in addition to the electron spin resonance
(ESR), N-\emph{V} centers have two very interesting internal degrees
of freedom: their narrow optical resonance might be used for coherently
converting microwave into optical quantum states of the field \cite{NVLukinPhotonEntanglement},
and their coupling to the nitrogen atom nuclear spin could give access
to coherence times much longer than with electron spins \cite{NVLukinNuclearSpinRegister}.

\begin{figure}[h]
\includegraphics[width=8cm]{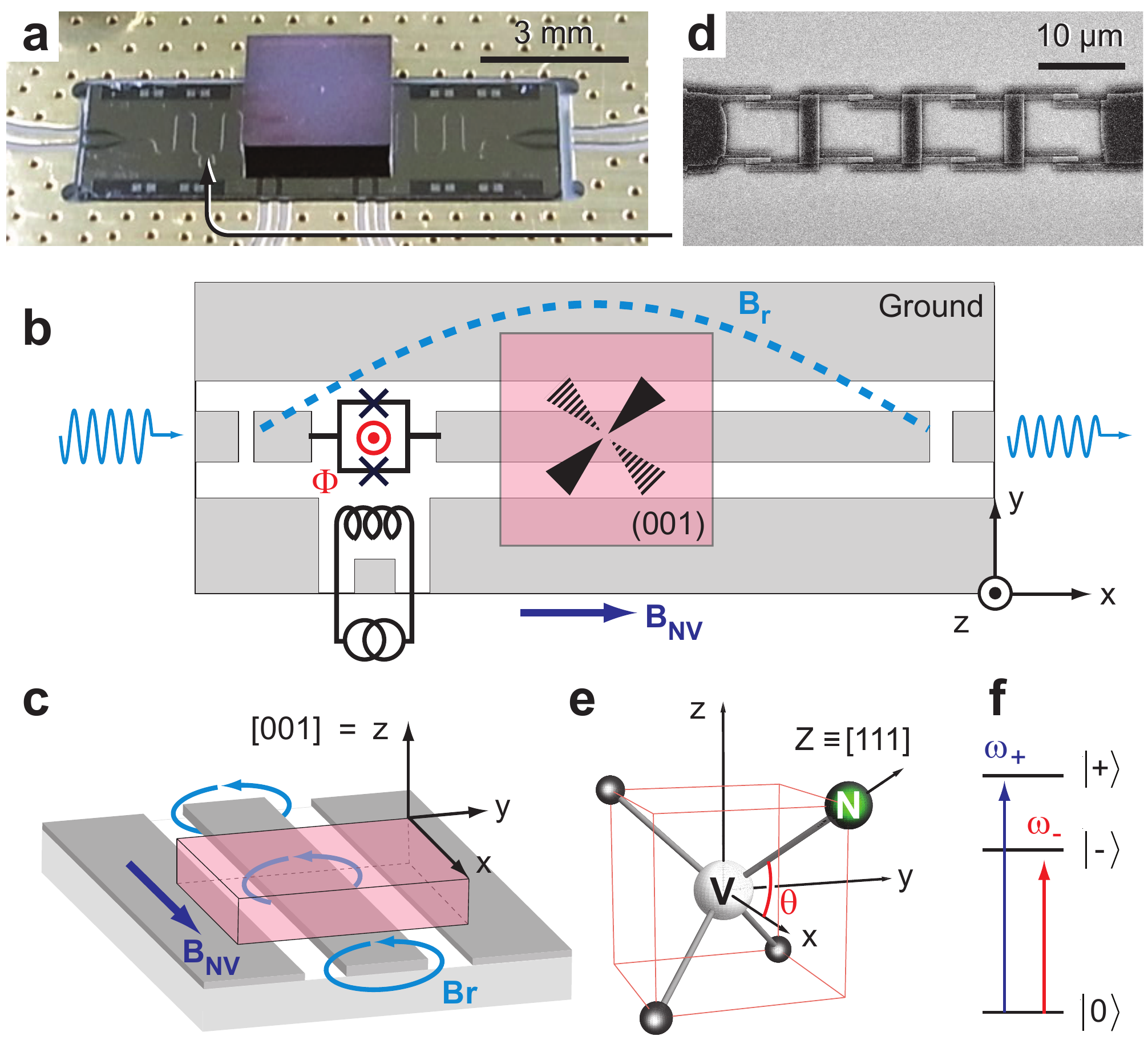}\caption{Scheme of the experiment. (a-c) Photograph, top and side view of the
sample : a diamond crystal is glued on top and in the middle of a
superconducting coplanar resonator with its $(001)$ surface facing
the chip. A SQUID array (d) is inserted in the central conductor to
tune the resonator frequency $\omega_{\mathrm{r}}(\Phi)$ with an
on-chip wire producing a flux $\Phi$ in the SQUIDs. A magnetic field
$\mathbf{B}_{\mathrm{NV}}$ is applied parallel to the $[100]$ direction.
(e) Sketch of a N-\emph{V} center with its vacancy (V) and nitrogen
atom (N), as well as three neighboring carbon atoms. Four equivalent
$<111>$ directions exist for the N-\emph{V} axis, all making the
same angle $\theta=55\text{\textdegree}$ with $[100]$, and thus
with $\mathbf{B}_{\mathrm{NV}}$. (f) Energy diagram of the N-\emph{V}
center, with the $\left|m_{\mathrm{S}}=0\right\rangle $ ground state
separated by $\omega_{\pm}$ from the excited states $\left|\pm\right\rangle $,
which are linear combinations of the pure spin states $\left|m_{\mathrm{S}}=+1\right\rangle $
and $\left|m_{\mathrm{S}}=-1\right\rangle $.}

\label{fig1}
\end{figure}

In this letter we report the observation of a vacuum Rabi splitting
\cite{VacuumRabiKimble,VacuumRabiRydberg,VacuumRabiCarmichael,VacuumRabiWeisbuch}
in the transmission spectrum of a superconducting resonator magnetically
coupled to an ensemble of $\sim10^{12}$ N-\emph{V} centers with a
collective coupling constant $g_{\mathrm{ens}}/2\pi=11$\ MHz. This
demonstrates the strong coupling of a microscopic spin ensemble to
a macroscopic circuit. In a parallel experiment, Schuster et al. \cite{Schuster}
observe a similar coupling to electronic spins in ruby and diamond
(see also \cite{Chiorescu}). The sketch of our experiment is shown
in Fig.\ \ref{fig1}. A diamond single-crystal ($3\times3\times0.5\mathrm{\, mm^{3}}$)
containing the N-\emph{V} centers is glued on top of a half-wavelength
niobium coplanar waveguide resonator, with a distance to the silicon
substrate less than $\sim0.5\mu\mathrm{m}$ to ensure a maximum spin-resonator
coupling. The diamond is positioned in the middle of the resonator
where the magnetic field is maximum, with its $(001)$ crystallographic
plane facing the chip. The spin Zeeman splitting can be tuned with
a magnetic field $\mathbf{B}_{\mathrm{NV}}$ parallel to the sample
surface along the $[100]$ axis within a few degrees. An array of
four superconducting quantum interference devices (SQUID) is inserted
in the resonator central conductor, away from the diamond crystal,
to make its frequency $\omega_{\mathrm{r}}(\Phi)$ tunable with the
magnetic flux $\Phi$ threading the SQUID loops \cite{TunableResonatorsPalacios}.
This flux is generated by passing current through an on-chip wire
so that the resonator can be brought in resonance with the spins without
changing their Zeeman splitting. The resonator transmission $\left|S_{21}\right|(\omega)$
is measured with a network analyzer, at powers low enough for the
current through the SQUID to stay well below its critical current,
so that the resonator behaves linearly \cite{TunableResonatorsPalacios}.

\begin{figure}
\includegraphics[width=8cm]{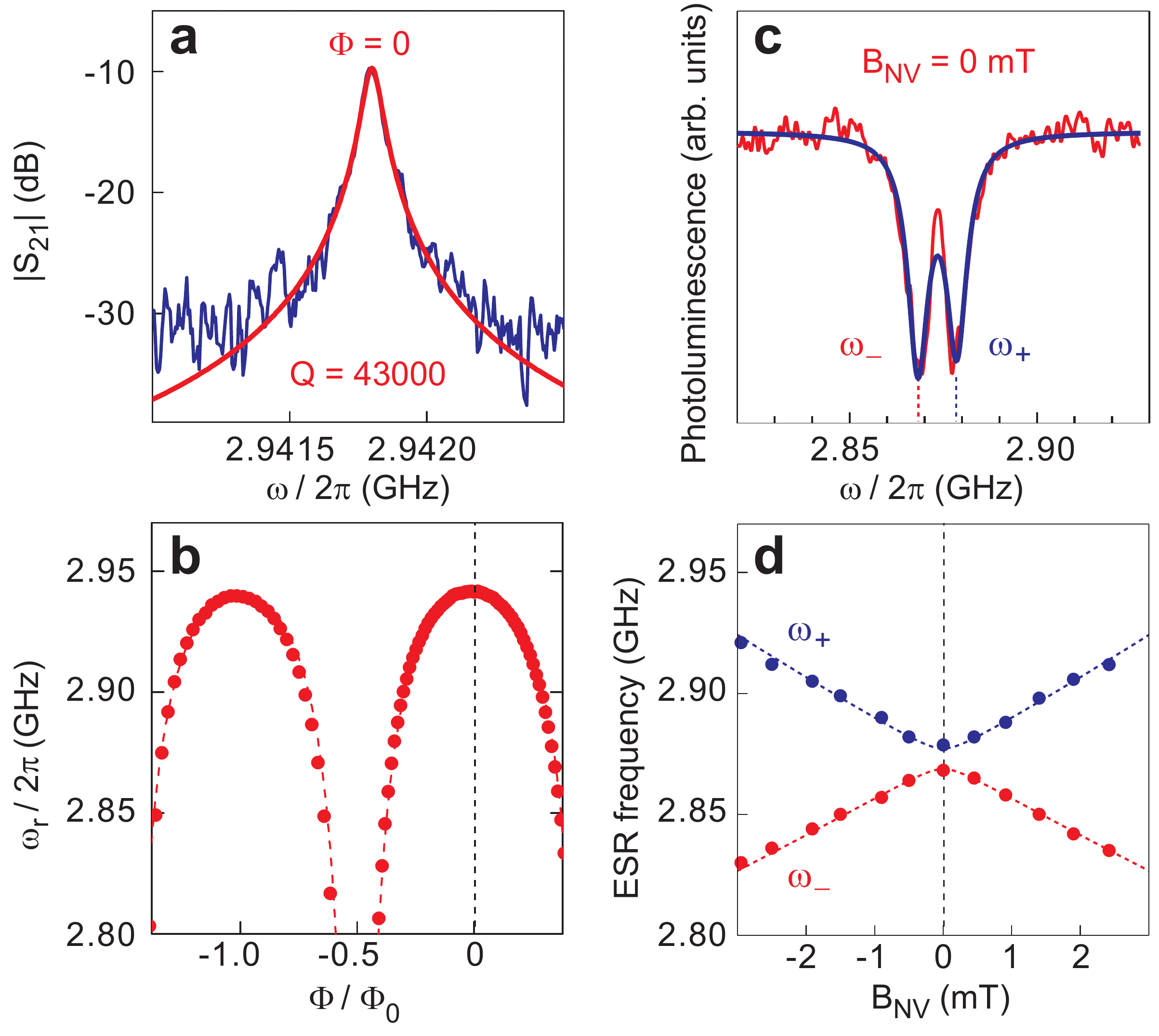}

\caption{(a) Resonance line of the coplanar resonator with no diamond on top,
measured at $40$~mK. (b) Measured (dots) and fitted (dashed line)
central frequency $\omega_{\mathrm{r}}$ as a function of the applied
flux $\Phi$, yielding a SQUID critical current $2I_{\mathrm{C0}}=2.2$~$\mu A$
and a bare resonator frequency $\omega_{\mathrm{r0}}/2\pi=3.012$~GHz.
(c) Optically detected ESR of the N-\emph{V} centers in the sample,
in zero magnetic field and at room-temperature. The $\omega_{\pm}$
lines are fitted by a sum of two $6$~MHz wide Lorentzians separated
by $10.4$~MHz. (d) Measured (dots) and fitted (dotted lines) dependence
of $\omega_{\pm}$ on $B_{\mathrm{NV}}$, yielding a zero-field splitting
$D/2\pi=2.873$~GHz and a strain-induced splitting $E/2\pi=4.3$~MHz.}

\label{fig2}
\end{figure}

We first characterize separately the coplanar resonator and the diamond
crystal. A typical spectrum of the resonator alone, cooled at $40$~mK,
is shown in Fig.\ \ref{fig2}a. The periodic dependence of the resonator
frequency with $\Phi$ is shown in Fig.\ \ref{fig2}b, in excellent
agreement with the model described in \cite{TunableResonatorsPalacios}
for parameters close to design values. On top of the modulation curve,
the quality factor $Q=5\cdot10^{4}$ results from the combination
of the coupling to the external line ($Q_{\mathrm{ext}}=10^{5}$)
and from internal losses ($Q_{\mathrm{int}}=10^{5}$). As already
observed in similar samples, $Q$ decreases when the resonator is
tuned to lower frequencies, possibly due to flux noise or losses in
the Josephson junctions \cite{TunableResonatorsPalacios}. Around
the N-\emph{V} centers frequency of $2.87$\ GHz, we find $Q\sim2\cdot10^{4}$
corresponding to a resonator energy damping rate $\kappa=\omega_{\mathrm{r}}/Q\sim0.9$
\ MHz.

\begin{figure*}[t]
\includegraphics[width=18cm]{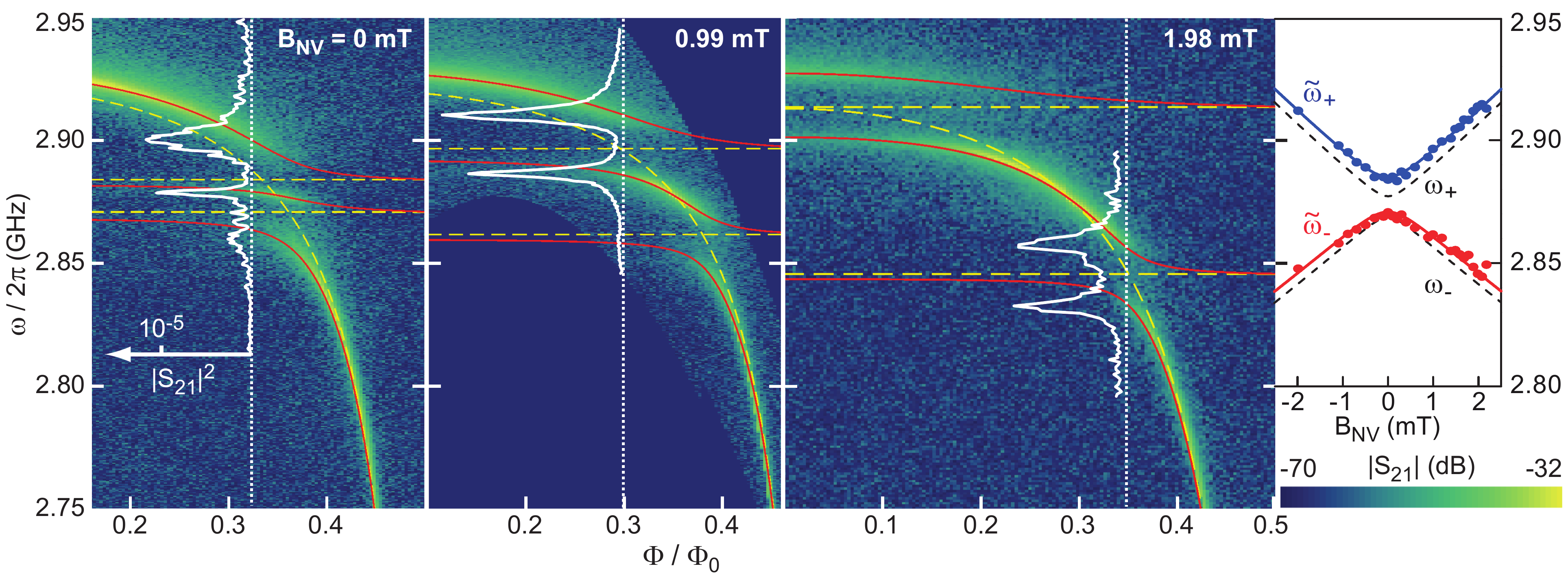}

\caption{Transmission $\left|S_{21}\right|$ of the resonator with diamond
on top (at $40$\ mK) as a function of $\Phi$ for $B_{\mathrm{NV}}=0,\,0.99$,
and $1.98$~mT. Two anticrossings are observed symmetrically around
$2.88$\ GHz, when the resonator frequency is resonant with the N-\emph{V}
transitions $\left|0\right\rangle \rightarrow\left|+\right\rangle $
and $\left|0\right\rangle \rightarrow\left|-\right\rangle $. Red
solid (yellow dashed) lines are fits to the eigenfrequencies of the
coupled (uncoupled) resonator-spins system as described in the text.
A transmission spectrum (white overlay) is also shown in linear units
in the middle of one of the two anticrossings, for each $B_{\mathrm{NV}}$
value. Right panel: Experimental frequencies $\tilde{\omega}_{\pm}$
(dots) resulting from the fits of $\left|S_{21}\right|(\omega,\Phi)$
for several $B_{\mathrm{NV}}$ values. The data points are then fitted
(solid lines) to the eigenfrequencies of Hamiltonian $H_{\mathrm{NV}}$
yielding $D/2\pi=2.878$\ GHz and $E/2\pi=7.2$\ MHz. For comparison,
the ESR frequencies $\omega_{\pm}(B_{\mathrm{NV}})$ measured at room-temperature
(see Fig.\ \ref{fig2}) are also shown as dashed lines.}

\label{fig3}
\end{figure*}

The diamond crystal (of the high pressure high temperature Ib type,
with a nominal $100$~ppm nitrogen concentration) was irradiated
with $2.5$~MeV protons at a dose of $5\cdot10^{16}\mathrm{cm^{-2}}$
in order to create vacancies, and subsequently annealed at $900$~$\lyxmathsym{\textdegree}$C
for $10$~h to form negatively charged N-\emph{V} centers. The resulting
N-\emph{V} distribution is homogeneous over a $30$~$\mathrm{\mu m}$
depth from the irradiated surface and thus over the spatial extension
of the resonator magnetic field (characteristic decay length of about
$10$~$\mathrm{\mu m}$). The N-\emph{V} concentration $\rho=(1.2\,\pm\,0.3)\times10^{6}$~$\mu\mathrm{m}^{-3}$
was measured by comparing the sample photoluminescence under a laser
excitation at $532$~nm to the photoluminescence of an individual
center in the same conditions. A typical optically detected ESR spectrum
is shown in Fig.\ \ref{fig2}c: two resonance lines are observed
even in zero magnetic field, due to residual strain in the crystal
that lifts the degeneracy of states $\left|m_{\mathrm{S}}=+1\right\rangle $
and $\left|m_{\mathrm{S}}=-1\right\rangle $, yielding new eigenstates
$\left|+\right\rangle $ and $\left|-\right\rangle $. The ESR frequencies
measured as a function of $B_{\mathrm{NV}}$ {[}see Fig.\ \ref{fig2}(d){]}
are in quantitative agreement with the expected transitions $\omega_{\pm}(B_{\mathrm{NV}})$
between the energy eigenstates of the spin Hamiltonian $H_{NV}=\hbar DS_{\mathrm{Z}}^{2}+\hbar E(S_{\mathrm{X}}^{2}-S_{\mathrm{Y}}^{2})+g_{\mathrm{NV}}\mu_{\mathrm{\mathrm{B}}}\mathbf{S}\cdot\mathbf{B}_{\mathrm{NV}}$
\cite{NVstrain}, with a zero-field splitting $D/2\pi=2.873$~GHz,
a strain-induced splitting $E/2\pi=4.3$~MHz, a N-\emph{V} Landé
factor $g_{\mathrm{NV}}=2$, and $\mu_{\mathrm{B}}/h=14$~MHz/mT.
Here, $Z$ denotes the N-\emph{V} axis {[}see Fig.\ \ref{fig1}(e){]},
while $X$ and $Y$ are the directions of nonaxial strain in the diamond
matrix. Note that only two lines are observed in our geometry because
the four possible N-\emph{V} crystalline orientations are all at the
same angle $\theta=55\text{\textdegree}$ with the $[100]$ direction
and thus with $\mathbf{B}_{\mathrm{NV}}$. The measured ESR lines
have a $\gamma/\pi\sim6$~MHz FWHM linewidth, compatible with the
expected broadening due to dipolar interactions with the neighboring
nitrogen electronic spins ($S=1/2$), given the nominal $100$~ppm
nitrogen concentration \cite{CoherenceTime}. 

We now present measurements of the resonator transmission with the
diamond crystal on top. The microwave power used corresponds to a
maximum intracavity energy of about $85$ photons at resonance and
is well below the value required to saturate the spins. Whereas optical
pumping is required to observe the ESR at room-temperature, cooling
down the sample to $40$~mK already provides a strong spin polarization
in $\left|m_{\mathrm{S}}=0\right\rangle $. Two-dimensional plots
of the transmission spectrum as a function of $\Phi$ are presented
in Fig.\ \ref{fig3} for several values of $B_{\mathrm{NV}}$, showing
two avoided crossings when the resonator is tuned through the N-\emph{V}
center ESR frequencies. To account for these observations, we model
the resonator-spins system \cite{VacuumRabiKimble} by the Hamiltonian
$H/\hbar=\omega_{r}(\Phi)a^{\dagger}a+\tilde{\omega}_{\mathrm{+}}b_{+}^{\dagger}b_{+}+\tilde{\omega}_{\mathrm{-}}b_{-}^{\dagger}b_{-}+(g_{+}a^{\dagger}b_{+}+g_{-}a^{\dagger}b_{-}+h.c)$,
whose justification is given below. It describes the coupling of the
resonator mode (annihilation and creation operators $a$ and $a^{\dagger}$)
to two spin-wave modes (annihilation and creation operators $b_{\pm}$
and $b_{\pm}^{\dagger}$) representing the two ESR resonances at frequencies
$\tilde{\omega}_{\pm}(B_{\mathrm{NV}})$, with coupling constants
$g_{\pm}$. The transmission spectrum $|S_{21}|(\omega)$ is fitted
for each $\Phi$ by a sum of Lorentzian peaks, whose central frequencies
are then fitted to the eigenfrequencies of $H$ (red solid lines in
Fig.\ \ref{fig3}). At zero field, this fit yields $g_{+}/2\pi=g_{-}/2\pi=11.0\pm0.5$~MHz
and a cooperativity parameter $C=g^{2}/(\kappa\gamma)\simeq27$, which
brings further confirmation that our experiment is in the strong coupling
regime. The $\tilde{\omega}_{\pm}(B_{\mathrm{NV}})$ data points obtained
and shown in Fig.\ \ref{fig3} agree very well with the eigenfrequencies
of the spin Hamiltonian $H_{\mathrm{NV}}$ for $E/2\pi=7.2$~MHz
and $D/2\pi=2.878$~GHz. This establishes that the anticrossings
observed are indeed due to the magnetic coupling between the coplanar
resonator and the N-\emph{V} centers. The small changes in $E$ and
$D$ with respect to the room-temperature values are likely due to
thermal contraction upon cooling \cite{NVTemperatureDependence}. 

The coupled oscillator model also predicts that the widths of the
two peaks at each anticrossing should be equal to the average of the
resonator and spin linewidth. While this is approximately true at
$B_{\mathrm{NV}}=1.98$~mT, where the two lines have comparable linewidths
of $\thicksim7$~MHz in the middle of the anticrossing, this is clearly
not the case at $B_{\mathrm{NV}}=0$~mT, where the line inbetween
the two anticrossings is much narrower than the other lines and narrower
than the coupled model predicts (see Fig.\ \ref{fig3}). A more refined
model is needed to understand this phenomenon. We also note that the
$1.5$~MHz FWHM resonator linewidth (corresponding to $\kappa=9.4$~MHz)
in the presence of the diamond far away from the anticrossings is
much larger than in Fig.\ \ref{fig2}a, indicating that this particular
diamond sample introduces losses of unknown origin.

We now justify our model for the interaction of the spin ensemble
with the resonator. A single N-\emph{V} center with spin operator
$\mathbf{S}_{\mathrm{k}}$ located at position $\mathbf{r}_{\mathrm{k}}$
is coupled to the resonator magnetic field $\mathbf{B}_{\mathrm{r}}(\mathbf{r}_{\mathrm{k}})$
by the Hamiltonian $H_{\mathrm{k}}=g_{\mathrm{NV}}\mu_{\mathrm{B}}\mathbf{B}_{\mathrm{r}}(\mathbf{r}_{\mathrm{k}})\cdot\mathbf{S}_{\mathrm{k}}$.
Here $\mathbf{B}_{\mathrm{r}}(\mathbf{r}_{\mathrm{k}})=\mathbf{\delta B_{0}}(\mathbf{r}_{\mathrm{k}})(a+a^{\dagger})$
with $\mathbf{\delta B_{0}}(\mathbf{r}_{\mathrm{k}})$ the rms vacuum
fluctuations of the magnetic field at $\mathbf{r}_{\mathrm{k}}$.
Restricting ourselves to one of the two $\left|0\right\rangle \rightarrow\left|\pm\right\rangle $
ESR transitions, this Hamiltonian can be put under a Jaynes-Cummings
form $H_{\mathrm{k}}=g_{\mathrm{k}}(\mathbf{r}_{\mathrm{k}})(a\sigma_{\mathrm{+,k}}+a^{\dagger}\sigma_{\mathrm{-,k}})$,
with $g_{\mathrm{k}}(\mathbf{r}_{\mathrm{k}})=(g_{\mathrm{NV}}\mu_{\mathrm{B}}/\hbar)\mathbf{\delta B_{0}}(\mathbf{r}_{\mathrm{k}})\cdot\left\langle 0\right|\mathbf{S}_{\mathrm{k}}\left|\pm\right\rangle $
and $\sigma_{\mathrm{+,k}}$ ($\sigma_{\mathrm{-,k}}$) the raising
(lowering) operator of the corresponding transition \cite{ForInstance}.
The coupling Hamiltonian of the ensemble of $N$ spins to the resonator
is $H_{\mathrm{ens}}=a\sum_{k}g_{\mathrm{k}}(\mathbf{r}_{\mathrm{k}})\sigma_{\mathrm{+,k}}+h.c$,
indicating that one excitation in the resonator mode is coupled to
a well-defined coherent superposition of spin excitations (a spin-wave)
\cite{MolmerHolographicQC,HybridsMolmer}. In the limit where the
number of system excitations is small compared to $N$, this spin-wave
behaves as a harmonic oscillator described by an annihilation operator
$b=\sum(g_{\mathrm{k}}(\mathbf{r}_{\mathrm{k}})/g_{\mathrm{ens}})\sigma_{\mathrm{-,k}}$,
and the Hamiltonian can be rewritten $H_{\mathrm{ens}}=g_{\mathrm{ens}}(ab^{\dagger}+a^{\dagger}b)$
with $g_{\mathrm{ens}}=(\int\rho d\mathbf{r}\left|g(\mathbf{r})\right|^{2})^{1/2}$.
For a homogeneous distribution $\rho$ and a resonator and crystal
of lengths $L$ and $l$, this collective coupling constant can be
expressed as $g_{\mathrm{ens}}=g_{\mathrm{NV}}\mu_{B}\sqrt{\eta\alpha\mu_{\mathrm{0}}\hbar\omega_{\mathrm{r}}(\Phi)\rho}/2\hbar$.
Here, $\eta=(1/L)\int_{(L-l)/2}^{(L+l)/2}\sin^{2}\frac{\pi x}{L}dx$
and $\alpha=\int\left|\mathbf{\delta B_{0}}(\mathbf{r})\right|^{2}\sin^{2}\varphi(\mathbf{r})d\mathbf{r}/\int\left|\mathbf{\delta B_{0}}(\mathbf{r})\right|^{2}d\mathbf{r}$
are dimensionless factors describing respectively the fraction of
the resonator mode volume occupied by the spins and their average
orientation with respect to the resonator microwave field, $\varphi(\mathbf{r})$
being the angle between the N-\emph{V} axis and $\mathbf{B}_{\mathrm{r}}(\mathbf{r})$.
We stress that apart from $\eta$ and $\alpha$, the ensemble coupling
constant only depends on the spins density. For our sample, we estimate
$\eta=0.29$ and $\alpha=0.81$, which combined with the measured
N-\emph{V} density yields $g_{\mathrm{ens}}/2\pi=11.6$\ MHz, in
agreement with the fitted value. This confirms that the spin ensemble
is highly polarized and thus not far from thermal equilibrium.

Although the strong coupling regime was reached in this sample, the
resonator and spin linewidths should be reduced by $1$ order of magnitude
in order to implement a quantum memory. This will require eliminating
the extra microwave losses caused by the diamond crystal which are
presently limiting the resonator linewidth. The spin resonance linewidth
on the other hand is believed to be limited by dipolar interactions
with neighboring nitrogen electronic spins; improving the nitrogen
to N-\emph{V} conversion rate would reduce the N-\emph{V} linewidth
while maintaining the large collective coupling constant measured
here. The very same setup would then allow one to investigate the
coherent oscillations between the spin ensemble and the resonator,
using a rapid tuning of the resonator frequency \cite{TunableResonatorsPalacios}
in and out of resonance with the spins. This would open the way to
the storage and retrieval of a given microwave field at the single-photon
level in the spin ensemble. 

In conclusion we have observed vacuum Rabi splittings in the transmission
spectrum of a superconducting coplanar resonator magnetically coupled
to an ensemble of N-\emph{V} centers, with a collective coupling constant
as large as $11$~MHz. The position of these anticrossings is in
excellent agreement with the electron spin resonances, and the measured
coupling constant is well understood. These results therefore constitute
an experimental evidence for the coherent coupling of a spin ensemble
to a superconducting circuit, an essential step towards the implementation
of more complex hybrid quantum circuits in which superconducting qubits,
electron and nuclear spins, microwave and optical photons would be
coherently coupled.

We gratefully thank M. Pomorski, N. Tranchant, G. Dantelle, F. Grosshans,
N.D. Lai, F. Treussart, J. Botsoa, F. Lainé, F. Carrel, G. Balasubramanian,
J. Meijer, S. Pezzagna, R. Kalish, D. Twitchen, D. Colson, and A.
Forget for help in the diamond sample preparation, P. Sénat, J.-C.
Tack, and P.-F. Orfila for technical help, and we acknowledge useful
discussions with M. Afzelius, J. Isoya and within the Quantronics
group. We acknowledge the support of European Contract MIDAS.


\begin{thebibliography}{29}
\bibitem{WinelandRMP} C.F. Roos \emph{et al.}, Phys. Rev. Lett. \textbf{92},
220402 (2004). 

\bibitem{QubitsSupra} J. Clarke and F. Wilhelm, Nature \textbf{453},
1031 (2008).

\bibitem{DiCarlo2qb} L. DiCarlo\emph{ et al.}, Nature \textbf{460},
240 (2009).

\bibitem{VacuumRabiKimble} M.G. Raizen\emph{ et al.}, Phys. Rev.
Lett. \textbf{63}, 240 (1989).

\bibitem{WallraffSqrtN} J.M.Fink\emph{ et al.}, Phys. Rev. Lett.
\textbf{103}, 083601 (2009).

\bibitem{VacuumRabiRydberg} Y. Kaluzny \emph{et al.}, Phys. Rev.
Lett. \textbf{51}, 1175 (1983).

\bibitem{HybridsZoller} P. Rabl \emph{et al.}, Phys. Rev. Lett. \textbf{97},
033003 (2006).

\bibitem{HybridsKurizki} D. Petrosyan \emph{et al.}, Phys. Rev. A
\textbf{79}, 040304(R) (2009).

\bibitem{HybridsSchmiedmayer} J. Verdu \emph{et al.}, Phys. Rev.
Lett \textbf{103}, 043603 (2009).

\bibitem{HybridsImamoglu} A. Imamoglu, Phys. Rev. Lett. \textbf{102},
083602 (2009).

\bibitem{HybridsMolmer} J.H. Wesenberg \emph{et al.}, Phys. Rev.
Lett. \textbf{103}, 070502 (2009).

\bibitem{LukinQMemory} M. Fleischhauer and M. D. Lukin, Phys. Rev.
Lett. \textbf{84}, 5094 (2000).

\bibitem{GisinQMemory} M. Afzelius \emph{et al.}, Phys. Rev. Lett.
\textbf{104}, 040503 (2010) ; C. Simon \emph{et al.}, Eur. Phys. J.
D \textbf{58}, 1 (2010).

\bibitem{BriggsEnsemble} H. Wu \emph{et al.}, arXiv:0908.0101.

\bibitem{MolmerHolographicQC} K. Tordrup, A. Negretti, K. Mølmer,
Phys. Rev. Lett. \textbf{101}, 040501 (2008).

\bibitem{NVFluxQLukin} D. Marcos\emph{ et al.}, arXiv:1001.4048;
J. Twamley and S.D. Barrett, Phys. Rev. B \textbf{81}, 241202 (2010).

\bibitem{NVJelezko} F. Jelezko \emph{et al.}, Phys. Rev. Lett. \textbf{92},
076401 (2004).

\bibitem{NVLongCoherence} G. Balasubramanian \emph{et al.}, Nature
Mat. \textbf{8}, 383 (2009).

\bibitem{NVLukinPhotonEntanglement} D. Englund \emph{et al.}, arXiv:1005.2204.

\bibitem{NVLukinNuclearSpinRegister} M.V. Gurudev-Dutt \emph{et al.},
Science \textbf{316}, 1312 (2007).

\bibitem{VacuumRabiCarmichael} Y. Zhu \emph{et al.}, Phys. Rev. Lett.
\textbf{64}, 2499 (1990).

\bibitem{VacuumRabiWeisbuch} C. Weisbuch, M. Nishioka, A. Ishikawa,
Y. Arakawa, Phys. Rev. Lett. \textbf{69}, 3314 (1992). 

\bibitem{Schuster} D. Schuster \emph{et al.}, arXiv:1006.0242.

\bibitem{Chiorescu} I. Chiorescu et al., Phys. Rev. B \textbf{82},
024413 (2010).

\bibitem{TunableResonatorsPalacios} A. Palacios-Laloy \emph{et al.},
J. Low Temp. Phys. \textbf{151}, 1034 (2008); M. Sandberg \emph{et
al.}, Appl. Phys. Lett. \textbf{92}, 203501 (2008).

\bibitem{NVstrain} P. Neumann \emph{et al.}, New J. Phys. \textbf{11},
013017 (2009).

\bibitem{CoherenceTime} J.A. van Wyk, et al., J. Phys. D \textbf{30},
1790 (1997).

\bibitem{NVTemperatureDependence} V.M. Acosta\emph{ et al.}, Phys.
Rev. Lett. \textbf{104}, 070801 (2010).

\bibitem{ForInstance} The field in the middle of our resonator and
$1\,\mu\mathrm{\mathrm{m}}$ above the surface is $\delta B_{0}=0.53\,$nT,
yielding $g/2\pi=10.5$\ Hz. 
\end{thebibliography}
\end{document}